# Thermoelectric PbTe-CdTe bulk nanocomposite

M. Szot[1,2], K. Dybko[1,2], A. Mycielski[1,3], A. Reszka[1], R. Minikayev[1], P. Dziawa[1], T. Story[1,2]

[1]Institute of Physics, Polish Academy of Sciences, Aleja Lotników 32/46, PL-02668 Warsaw, Poland

[2]International Research Centre MagTop, Institute of Physics, Polish Academy of Sciences, Aleja Lotników 32/46, PL-02668 Warsaw, Poland

[3]PUREMAT Technologies, 32/46, PL-02668 Warsaw, Poland

## Abstract

The preparation method of thermoelectric PbTe-CdTe semiconductor nanocomposite in the form of a bulk material doped with Bi, I or Na, intended for production the mid-temperature thermoelectric energy generators is presented. The method takes advantage of the extremely low mutual solubility of both semiconductors, resulting from their different crystal structure, and is based on a specifically designed Bridgman growth procedure. It is shown that the formation of zinc-blende crystalline CdTe grains in the rock-salt matrix of thermoelectric PbTe can be forced during the synthesis of a composite by introducing Cd in the form of CdTe compound and choosing the growth temperature above the melting point of PbTe but below the melting point of CdTe. X-ray diffraction and SEM-EDX spectroscopy analyzes as well as basic electric and thermoelectric characterization of the nanocomposite samples containing 2, 5 and 10 at. % of Cd showed that using proposed growth procedure, it is possible to obtain both *n*-type (Bi- or I-doped) and *p*-type (Na-doped) material with carrier concentration of $1 \div 5 \times 10^{19}$ cm$^{-3}$ and uniformly distributed CdTe grains with a diameter of the order of 100 nm.

## Introduction

Thermoelectric generators and coolers are simple solid-state devices employing the ability of electrons to simultaneously transfer heat and electric charge. They operate silently and without failure for many years because the only moving elements of these devices are electrons. Thus, thermoelectric devices with sufficiently high efficiency could became an alternative for other unconventional methods of large-scale energy managing and harvesting [1]. As follows directly from the formula for thermoelectric dimensionless figure of merit parameter $ZT = (\sigma S^2/\kappa)\, T$, good thermoelectric material should be characterized by high electric conductivity $\sigma$ and high Seebeck coefficient $S$ (thermoelectric power) but by low thermal conductivity $\kappa$ at given temperature $T$. As opposed to conventional method of improving the thermoelectric efficiency of known materials by proper doping or alloying with other elements, nanostructuring is one of new promising strategies in this regard. Nanostructuring implies preparation of low-dimensional structures basing on good thermoelectrics or introducing into them nano- and micro-grains of the same or other material phases. It is anticipated that reducing the dimensionality of thermoelectrics may enhance the thermoelectric power $S$ due to "sharpened" density of states of charge carriers and simultaneously decrease the thermal conductivity $\kappa$ by phonon scattering on interfaces in the structure [2,3]. Alloying thermoelectric materials with other elements or incorporation into them of grain or other phases is associated with further suppression of its thermal conductivity due to alloy disorder and grain boundaries scattering of phonons [4].

PbTe is well-known thermoelectric material exhibiting the best performance at temperatures around 600÷800 K. This narrow gap semiconductor together with wide gap CdTe constitute an ideal materials platform for researching both conventional and new strategies of improving $ZT$ since bulk, monocrystalline ternary solid solution [5,6], bulk polycrystalline materials [7-11] as well as the low dimensional structures can be obtained with these semiconductors as component materials [12-14]. In particular, it was expected that small admixture of Cd to PbTe may result in the appearance of an additional density of states at the Fermi level as it was shown in the case of PbTe:Tl [15]. However, rock-salt PbTe and zinc-blende CdTe exhibit very limited mutual solubility due to their different crystal structures [16]. For this reason, if applied rapid quenching [11] or hot-pressing growth methods [10], usually a polycrystalline material with undesirable CdTe precipitation embedded in low Cd content $Pb_{1-x}Cd_xTe$ matrix (up to 3 at.% of Cd) is obtained. To produce monocrystalline bulk $Pb_{1-x}Cd_xTe$ material with higher Cd content (reaching 10 at.% of Cd) sophisticated self-selected vapor growth (SSVG) method at lower temperature must be used [5,6,17-20]. In the case of monocrystalline $Pb_{1-x}Cd_xTe$ (x=0.1) samples addition of Cd results in twofold decrease of thermal conductivity and increase of Seebeck coefficient owing to rising the energy gap and effective mass of holes (stemming from increasing participation of Σ-sub-band heavy holes in transport phenomena) [6]. However, benefit from this is strongly limited for samples with Cd content higher than 2 at. % due to the reduction of mobility of current carriers, being a consequence of alloy scattering. Even for a PbTe-CdTe composite being an effect of non-intentional formation of CdTe grains in PbTe, relatively high *ZT* parameter, resulting, among others, from decreased thermal conductivity caused by scattering of phonons on PbTe-CdTe interfaces, was observed [9-11]. Thus, to fully use the thermoelectric potential of PbTe-CdTe material system, intentional preparation of composite with controlled distribution of CdTe precipitations in PbTe matrix seems to be the best strategy. The method of preparation of low-dimentional layered PbTe-CdTe nanocomposite with well controllable distances as well as spatial sizes of CdTe grains in PbTe matrix using molecular beam epitaxy technique (MBE) has been explored successfully previously in the context of both thermoelectric [12,13] and optical use [14]. In particular, for such a composite an increase of the Seebeck coefficient by about 30 % relative to pure PbTe was observed due to the sharpening of the density of states in the area between the CdTe grains [12]. However, use of low-dimentional structures in real applications is definitely limited by the small working volume and expensive methods of growing such structures. Therefore, nanostructured bulk materials are expected to be a real breakthrough in significantly increasing the efficiency of thermoelectric devices.

In this paper, we demonstrate the modified protocol of growth by the Bridgman method protected by our European patent [21], which we have used to obtain a bulk crystalline PbTe-CdTe nanocomposite i.e. a material useful from the application point of view. The method we used was successfully utilized to grow high quality II-VI and IV-VI semiconductor compounds for thermoelectric [22] and X-ray detection applications [23-27]. In contrast to the SSVG method used by us to growth of monocrystalline $Pb_{1-x}Cd_xTe$ solid solution, material obtained by the Bridgman technique is characterized by polycrystalline structure what is a desired effect from thermoelectrical point of view. In presented growth protocol we deliberately use the properties of PbTe-CdTe heterosystem to incorporate CdTe into PbTe matrix in the form of CdTe precipitations by properly chosen growth regime.

**Growth method and experiment**

The input material for obtaining the PbTe-CdTe composite was prepared in the form of pure Pb and Te elements as well as CdTe in compound form. Lead of purity of N5 (99.999%) was provided as a 6 mm diameter rod cut into pieces. Tellurium and compound cadmium telluride, both of purity of N6, were provided in the form of chipping. Te and CdTe were crushed in mortar into grains with diameter of c.a. 1 mm and mixed with Pb, which is not prone to crumbling. The components sealed in quartz ampule with diameter of 20 mm, evacuated to the pressure of $10^{-6}$ Torr, were placed in upper part of the two-zone electric furnace with a special design that allows it to be rocked during the growth process [21, 28]. Next, upper zone of the furnace was heated to 950 °C for 48 hours - i.e. to temperature 33 °C above the melting point of PbTe (917 °C [29,30]) but significantly below the melting point of CdTe (1092 °C [31]). This selection of the input form of the components and the growth temperature prevented the dissolution of the CdTe compound in PbTe. The temperature of the lower zone of the furnace during the growth process was kept at 850 °C i.e. below the melting point of PbTe. After complete dissolution of input material at temperature 950 °C, to obtain homogenized mixture of PbTe with uniformly distributed CdTe, the furnace was rocked from a vertical to a horizontal position with frequency of 0.5 Hz for 15 minutes. After that, the furnace was locked in its upright position and the ampoule was quickly (within 5 sec.) abandoned in it so that the part of the ampoule with the dissolved material was below the crystallization point of PbTe and the upper, empty part of the ampoule, was still above the crystallization point of PbTe. Keeping the upper part of the ampoule at an elevated temperature protects the rapidly cooled material against defects such as craters and voids. In turn, rapid cooling of the material after mixing by rocking the furnace is to ensure a uniform distribution of CdTe precipitations over the entire volume of the final composite material. The ampoule was kept in the down position for 3-4 hours and finally the furnace was turned off to cool the crystal to room temperature over 10 hours.

After the growth process the ingots shown in Fig. 1(a) were cut with a fine wire saw into 2-mm slices enabling further structural and electrical analysis of the composite obtained. The chemical composition, crystal quality and uniformity of CdTe phase distribution along the composite ingots (Fig. 1(b)) have been checked by XRD, SEM and EDX measurements. XRD data were analyzed by the Rietveld method using the FullProf2k program. In the initial model of the diffraction pattern, the background was set manually, the shape function of the Pearson VII peak profile was adopted and a set of appropriate refined parameters were applied. Electrical parameters were determined by Hall effect and conductivity measurements using standard six-contact Hall bar geometry. Finally, the thermoelectric figure of merit parameter *ZT* was determined directly using Harman method with radiation correction [22,32-34].

**Experimental results and discussion**

X-ray diffraction measurements of the PbTe-CdTe nanocomposite grown by proposed by us method revealed that they are polycrystalline with single $Pb_{1-x}Cd_xTe$ rock-salt phase in the case of samples with nominal Cd content up to 2 at. % while for samples with higher nominal Cd composition additional CdTe zinc-blende phase is observed. This conclusion was drawn on the basis of the change in the crystal lattice parameter and the evident asymmetry of the (111), (311), and (331) (Pb,Cd)Te reflexes in the XRD spectra of the PbTe-CdTe composite with the same Miller indexes as the main peaks of CdTe – see Figs. 2 and 3. Figure 3 (a) shows the dependence of the lattice parameter $a_0$ of the PbTe-CdTe composite on Cd content in the sample. It is clearly visible that lattice parameter of the composite follows the changes of $a_0$ observed for $Pb_{1-x}Cd_xTe$ monocrystalline samples obtained by SSVG method up to *x* of about 0.02 with *da/dx* = - 0.43 Å [5]. For samples with Cd content between 2 and 7 at. % lattice parameter changes significantly slower with *da/dx*= - 0.28 Å. For higher *x* it remains almost unchanged with values varying around that observed for 1-2 at. % samples, i.e. 6.46 Å. Such behavior was expected and is in agreement with previous reports relating to the PbTe bulk materials with addition of Cd obtained by rapid quenching or hot pressing methods [10,11]. More careful analysis of XRD data indicates that Cd ions not incorporated to the PbTe crystal lattice are present in the composite in the form of a CdTe precipitates. This manifests itself in the asymmetry of XRD reflexes for samples with nominal Cd content higher than 2 at. %. The main peaks of CdTe are partially overlapped with those peaks of $Pb_{1-x}Cd_xTe$ phase, what is shown in the example of (111) peak in Figure 3. Deconvolution of (111) reflexes for nominally 5 and 10 at. % samples presented in Fig. 4 (b) and 4(c) allows to determine of the CdTe content in these samples at the level of 2 and 6 % of the total molar mass, respectively. In the case of 2 % - sample, all introduced Cd ions seems to be dissolved in the PbTe matrix. Conclusions from the XRD data analysis were confirmed by the results of SEM and EDX measurements made for the PbTe-CdTe composite obtained by our method. As can be seen in Fig. 5 (a), in the case of a nanocomposite sample with nominal 2 at. % addition of Cd, according to the expectations, no precipitates are visible in its SEM image. However, as shown in Figs. 5 (b) and 5 (c), for 5% - and 10% samples precipitates with the diameter of the order of 100 nm and 300 nm respectively, uniformly distributed in the host matrix are clearly visible. EDX analysis of samples, shown in Fig. 6 for 10% case, revealed that they are CdTe precipitates embedded in the $Pb_{1-x}Cd_xTe$ matrix with a Cd content of about 1.5 at. %. The dependency of the CdTe grain diameter in nanocomposite on nominal Cd content is quite understood if take into account two times higher amount of Cd incorporated to the nanocomposite for 10%-sample and similar fraction of Cd dissolved in matrix in the case of both 5%- and 10% samples. Further, visible in Fig. 5 (d) arrangement of CdTe precipitates in the form of equidistant layers of evenly distributed grains suggest presence of some kind of self-organization process during growth of the bulk PbTe-CdTe nanocomposite. It seems to be characteristic for PbTe-CdTe heterosystem and was observed by us also in the case of layered PbTe-CdTe samples grown by molecular beam epitaxy technique (MBE) [12]. However, for layered composite the equidistance of layers of CdTe grains is forced by arrangement of the sample structure along growth direction i.e. by the initial thicknesses of the PbTe and CdTe layers while the organization process of the CdTe grains in plane of the sample is autonomous. In the case of the bulk nanocomposite the self-organization of the CdTe grains in whole volume of the material most likely results from the stress field inside the composite during the crystallization process. This observation, together with the dependence of the grain diameter on the CdTe composition in the material, holds great promise for thermal conductivity engineering through size-dependent phonon scattering at grain boundaries.

In order to confirm the possibility of controlling the type of electrical conductivity and carrier density of the final PbTe-CdTe nanocomposite, the crystals were additionally doped with sodium for *p*-type and with bismuth or iodine for *n*-type conductivity. While undoped samples have *p*-type conductivity with hole concentrations not exceeding $2 \times 10^{18}$ cm$^{-3}$, intentionally doped PbTe-CdTe composite samples exhibit both types of conductivity with carrier density and mobility at room temperature varying in the range $1\div5 \times 10^{19}$ cm$^{-3}$ and $200\div1100$ cm$^2$V$^{-1}$s$^{-1}$, respectively, as it is shown in Tab. 1. Figure 7 shows examples of the dependence of the thermoelectric figure of merit parameter *ZT* on temperature measured directly by the Harman method for 2% composite doped with iodine (black points) or bismuth (blue points) for *n*-type, and doped with sodium (red points) for *p*-type electric conductivity. As can be seen in Fig. 7, the *ZT* parameter increases with temperature for all presented composites reaching the value of 0.7÷0.8 at temperatures around 650 K what is very good result for not optimized electrically samples. Optimization of carrier density by appropriate doping significantly improve *ZT* as it follows from our theoretical predictions for PbTe-CdTe crystals [6].

**Summary**

We prepared the polycrystalline bulk PbTe-CdTe nanocomposite using a special protocol of modified Bridgman method. Of particular importance for our method is the choice of growth step at temperature window below the melting point of CdTe and the incorporation of Cd to the process in the form of CdTe compound what forces the formation of CdTe grains in PbTe. The obtained composite is characterized by an uniform distribution of CdTe crystallites throughout its volume, resulting from the rocking of the furnace during the synthesis of the material. The advantage of the proposed method is the self-limiting solubility of both semiconductors, which allows to control the diameter of the CdTe grains and the amount of Cd embedded in the PbTe crystal lattice. In addition, unlike SSVG growth of monocrystals, using the modified Bridgman method, additional doping with other elements such as Bi, I or Na is possible, what is very important for controlling the type of conductivity and optimizing electrical parameters to maximize the *ZT* figure of merit parameter of the final composite. Since the presented method is based on the basic properties of the material and requires only standard equipment, it can be easily implemented in the industry to obtain a PbTe-CdTe composite in the form of a bulk material with an easy scalable size, ready for the direct preparation of thermoelectric devices.

**Acknowledgments**

The research was partially supported by the National Centre for Research and Development (Poland) through grant TERMOD TECHMATSTRATEG2/408569/NCBR/2019, by the National Science Centre through Grant No. 2021/41/B/ST3/03651 and by the Foundation for Polish Science through the IRA Programme co-financed by EU within SG OP (Grant No. MAB/2017/1)

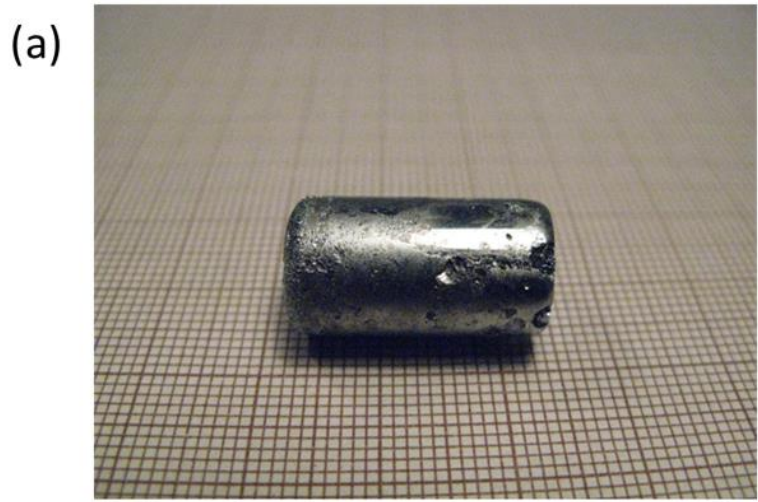

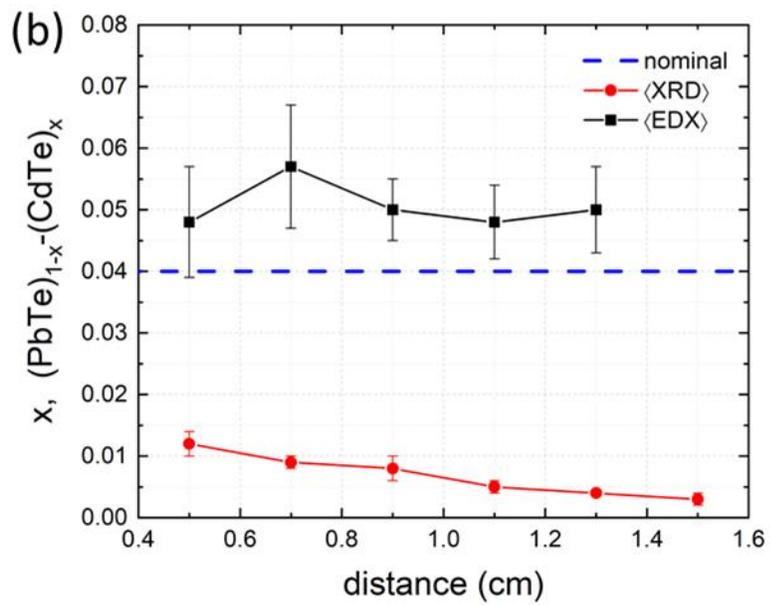

Fig. 1. a) $(PbTe)_{1-x}$-$(CdTe)_x$ nanocomposite ingot obtained by modified Bridgman method. b) CdTe phase distribution along the ingots of nominal $(PbTe)_{0.96}$ - $(CdTe)_{0.04}$ composite measured by XRD and EDX techniques.

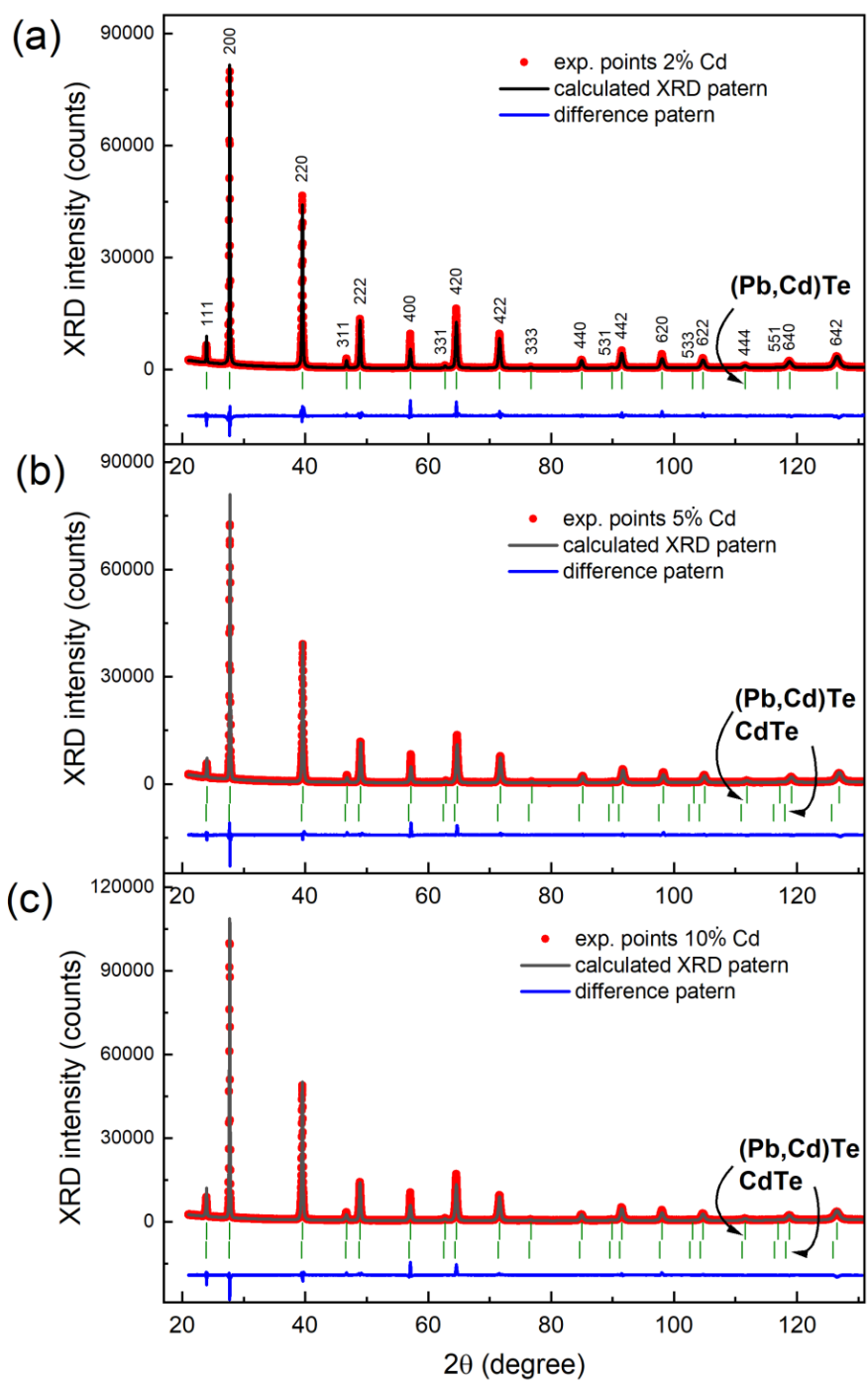

Fig. 2. The results of Rietveld refinement of XRD spectra for PbTe-CdTe samples with nominal Cd content of 2 at. % (a), 5 at. % (b) and 10 at. % (c). The experimental points are indicated by dots; the calculated patterns by the solid line; The positions of Bragg reflections of each phases by short vertical bars. At the bottom of each subfigure the difference patterns are displayed.

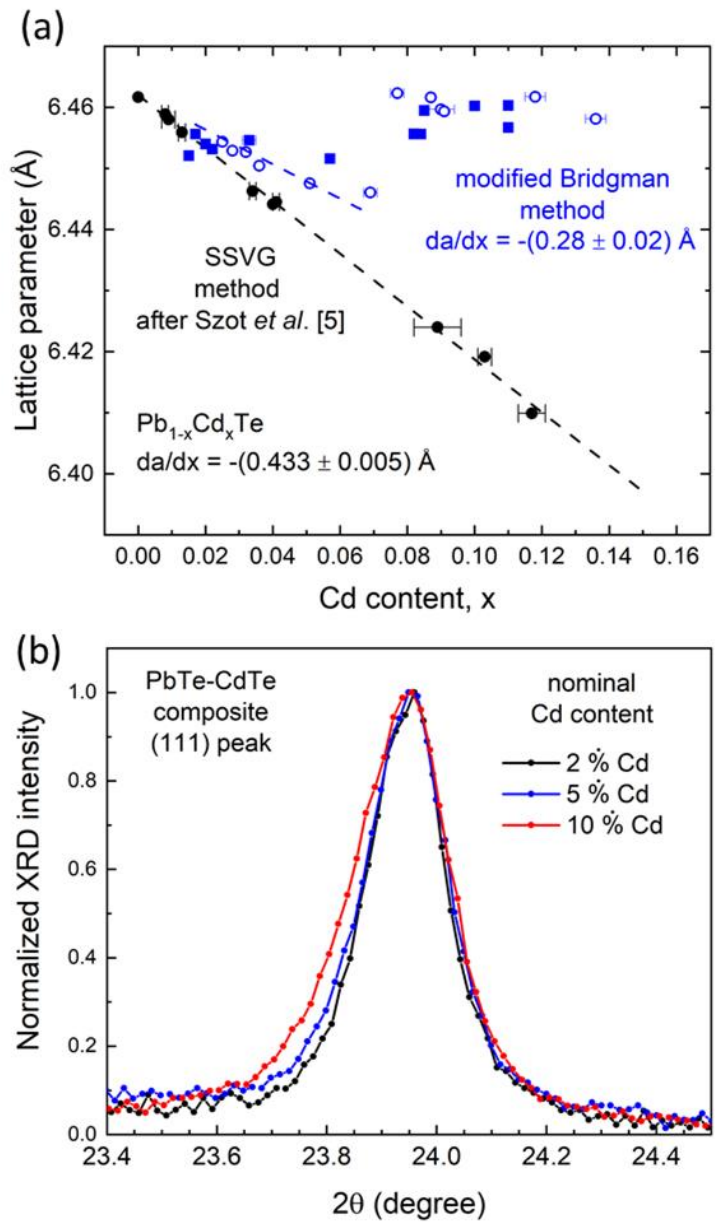

Fig. 3. a) Lattice parameter $a_0$ (XRD measurements) of the PbTe-CdTe nanocomposite (blue points) as a function of Cd content determined by EDX analysis. Black points – reference data for monocrystalline material grown by SSVG method [5]. b) The comparison of the asymmetry of the XRD (111) Bragg peaks for different samples of the PbTe-CdTe composite.

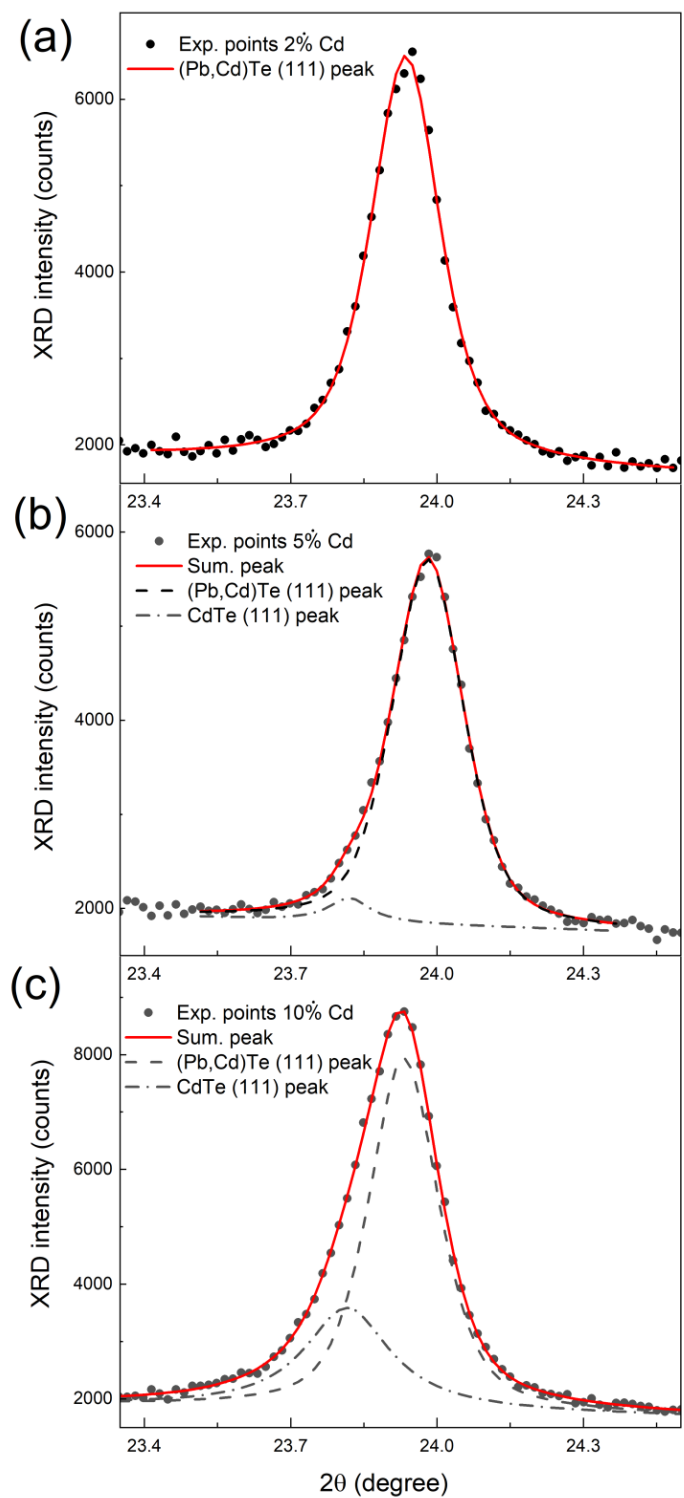

Fig. 4 Deconvolution of the (111) Bragg peaks for 2 % (a), 5 % (b) and 10 % (c) of Cd content.

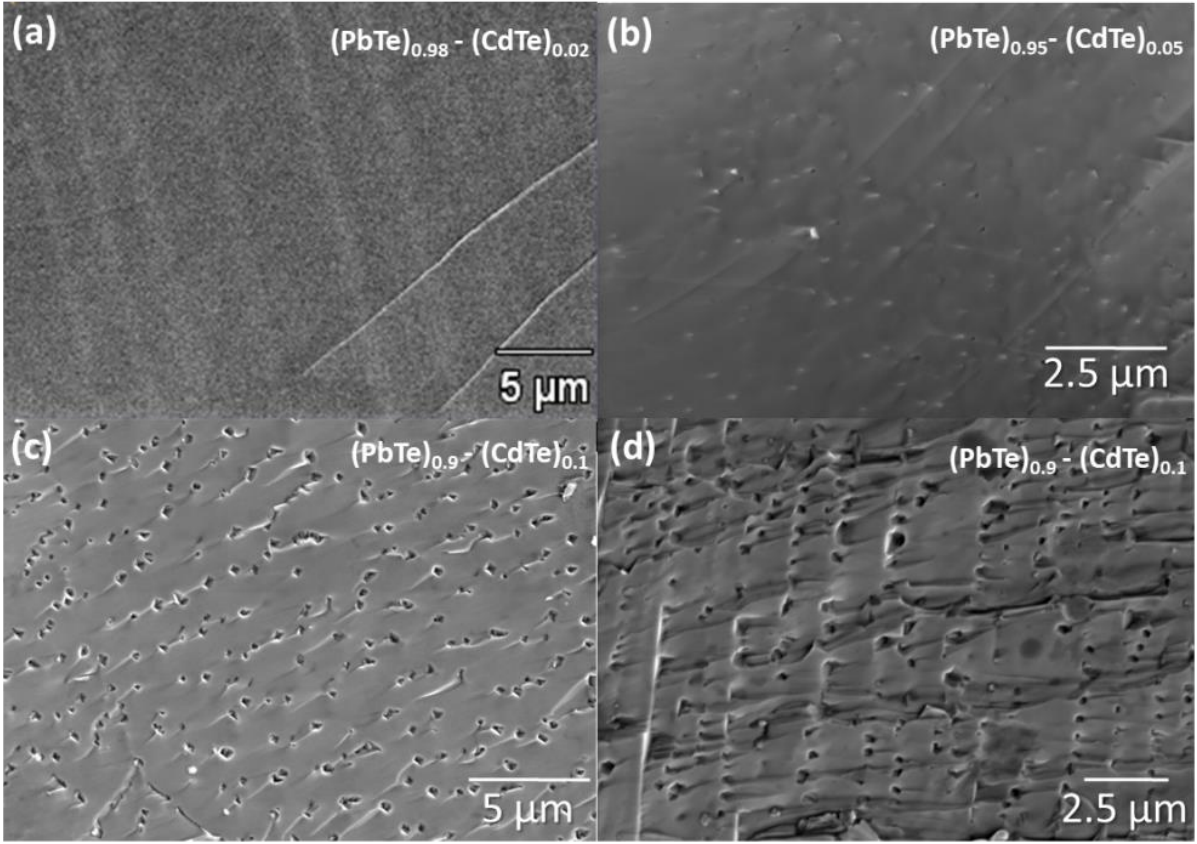

Fig. 5. SEM images of $(PbTe)_{1-x}$-$(CdTe)_x$ nanocomposite for different nominal Cd content $x$ = 0.02 (a), $x$ = 0.05 (b), two different places of sample with $x$ = 0.1 - (c) and (d).

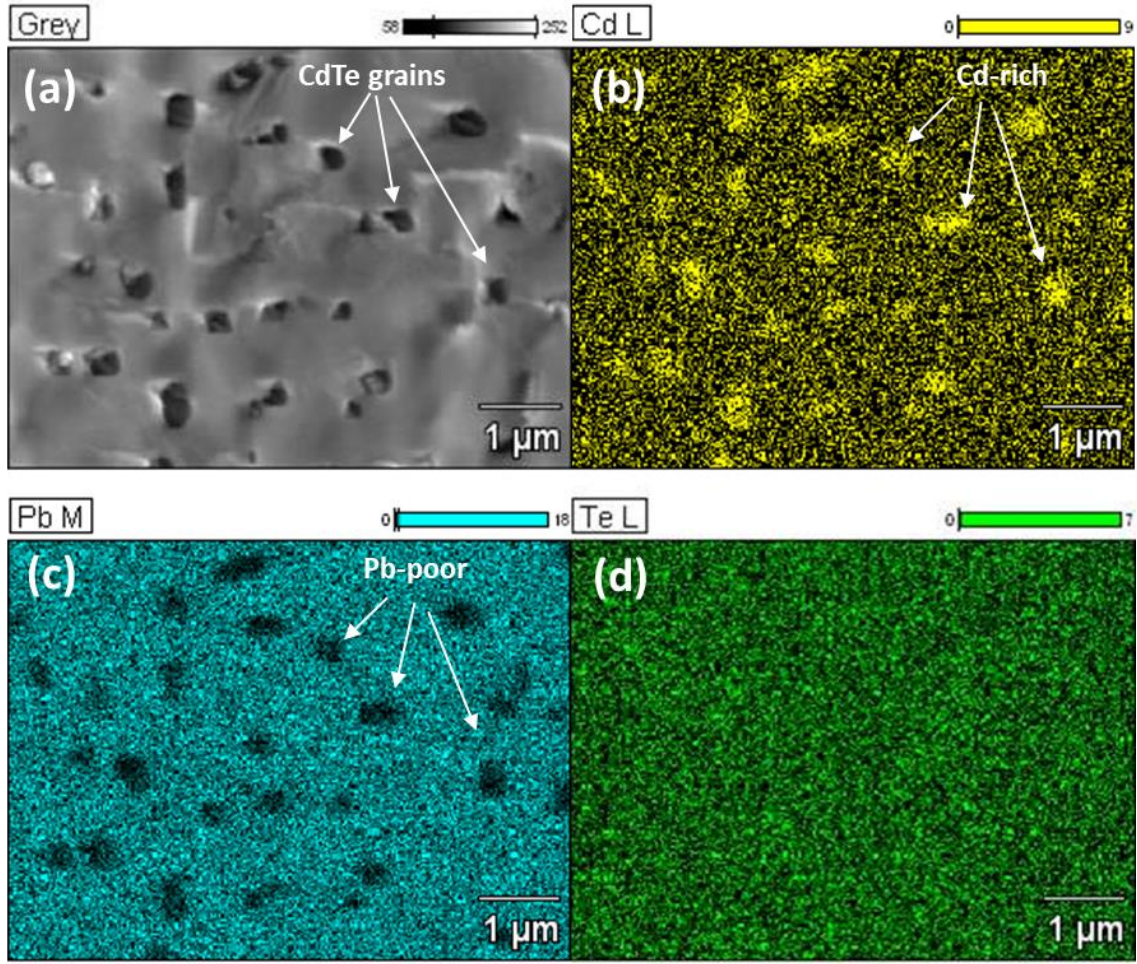

Fig. 6. (a) SEM images of (PbTe)$_{0.9}$-(CdTe)$_{0.1}$ nanocomposite sample. Distribution of Cd (b), Pb (c) and Te (d) elements collected with EDX technique.

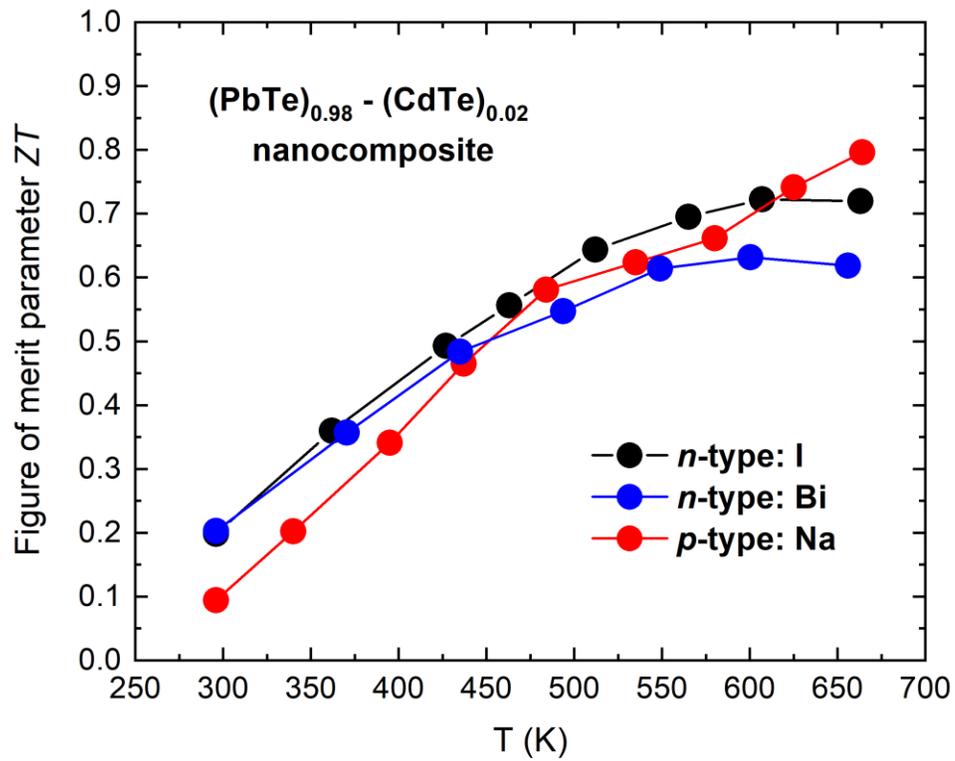

Fig. 7. Temperature dependence of the thermoelectric figure of merit parameter *ZT* of $(PbTe)_{0.98}$ - $(CdTe)_{0.02}$ composites doped with iodine, bismuth (*n*-type) or sodium (*p*-type) versus temperature measured by the Harman method (with a radiation correction).

| Cd nomin. content [at. %] | dopant | type of electrical conductivity | carrier density [cm$^{-3}$] | carrier mobility [cm$^2$V$^{-1}$s$^{-1}$] | conductivity [Ω$^{-1}$cm$^{-1}$] |
|---|---|---|---|---|---|
| 1 | - | p | 2.2 10$^{18}$ | 550 | 193 |
| 1 | - | p | 1.8 10$^{18}$ | 535 | 158 |
| 2 | Na | p | 4.8 10$^{19}$ | 172 | 1128 |
| 2 | I | n | 1.26 10$^{19}$ | 905 | 1824 |
| 2 | Bi | n | 1 10$^{19}$ | 1100 | 1900 |
| 4 | Na | p | 1.8 10$^{19}$ | 280 | 785 |
| 5 | - | p | 1.1 10$^{19}$ | 634 | 1295 |
| 10 | - | p | 3.85 10$^{17}$ | 260 | 16 |

Tab. 1. Room temperature electrical parameters for selected doped and undoped PbTe-CdTe nanocomposite with different nominal Cd content.